\documentclass[12pt, preprint]{aastex}

\def\ion[#1 #2]{#1\,{\sc #2}}

\begin{document}

\title{Chromospheric Evaporation in an M1.8 Flare Observed by the
  Extreme-ultraviolet Imaging Spectrometer on {\it Hinode}}

\author{G. A. Doschek \& H. P. Warren} \affil{Space Science Division, Naval Research
Laboratory, Washington, DC 20375, USA}

\author{P. R. Young} \affil{College of Science, George Mason
University, 4400 University Drive, Fairfax, VA 22030}

\begin{abstract}

We discuss observations of chromospheric evaporation for a complex
flare that occurred on 9 March 2012 near 03:30 UT obtained from the
Extreme-ultraviolet Imaging Spectrometer (EIS) on the {\it Hinode}
spacecraft.  This was a multiple event with a strong energy input that
reached the M1.8 class when observed by EIS.  EIS was in raster mode
and fortunately the slit was almost at the exact location of a
significant energy input.  Also, EIS obtained a full-CCD spectrum of
the flare, i.e., the entire CCD was readout so that data were obtained
for about the 500 lines identified in the EIS wavelength ranges.
Chromospheric evaporation characterized by 150-200 km/s upflows was
observed in multiple locations in multi-million degree spectral lines
of flare ions such as \ion[Fe xxii], \ion[Fe xxiii], and \ion[Fe
xxiv], with simultaneous 20-60 km/s upflows in million degree coronal
lines from ions such as \ion[Fe xii]-\ion[Fe xvi].  The behavior of
cooler, transition region ions such as \ion[O vi], \ion[Fe viii],
\ion[He ii], and \ion[Fe x] is more complex, but upflows were also
observed in \ion[Fe viii] and \ion[Fe x] lines.  At a point close to
strong energy input in space and time, the flare ions \ion[Fe xxii],
\ion[Fe xxiii], and \ion[Fe xxiv] reveal an isothermal source with a
temperature close to 14 MK and no strong blueshifted components.  At
this location there is a strong downflow in cooler active region lines
from ions such as \ion[Fe xiii] and \ion[Fe xiv].  We speculate that
this downflow may be evidence of the downward shock produced by
reconnection in the current sheet seen in MHD simulations.  A sunquake
also occurred near this location.  Electron densities were obtained
from density sensitive lines ratios from \ion[Fe xiii] and \ion[Fe
xiv].  The results are combined with context Atmospheric Imaging
Assembly (AIA) observations from the {\it Solar Dynamics Observatory}
to obtain an overview of the flare and data are presented that can be
used for predictive tests of models of chromospheric evaporation as
envisaged in the Standard Flare Model.

\end{abstract}


\section{INTRODUCTION}

Understanding energy release and transport in solar flares is one of
the outstanding problems in contemporary solar physics.  Solar flares
have been observed from the radio to gamma ray wavelength ranges and
solar flare models have been developed and continue to evolve and
reach new levels of sophistication. The prevailing current view is
embodied in the so-called Standard Flare Model, which has its origins
in models such as proposed by Carmichael (1964), Sturrock (1968),
Hirayama (1974), and Kopp \& Pneuman (1976). In the Standard Model,
magnetic reconnection occurs in a current sheet in the corona, and
results in plasma heating and particle acceleration.  Heat conduction
and high energy particles propagate along magnetic flux lines down
into the chromosphere, heating the chromosphere and driving it upwards
at multi-million degree temperatures into magnetic flux tubes closed
by the reconnection process.  The evaporated plasma explains the
intensity and high electron density of soft X-ray emitting coronal
loops.  In addition, models of particle deposition into the
chromosphere have shown that in some cases there is also a downward
recoil produced by momentum conservation (e.g., Canfield et al. 1987;
Fisher et al. 1985).  Until recently, details of the evaporation
process have been quite limited observationally.

The first spectroscopic observation of multi-million degree
evaporation during flares was reported by Doschek et al. (1980) and
Feldman et al. (1980).  These authors noticed a blue spectral wing
component present on the resonance lines of \ion[Ca xix] near 3.17
\AA\ and \ion[Fe xxv] near 1.85 \AA.  The spectra were from the
SOLFLEX spectrometers on the Department of Defence (DoD) {\it P78-1}
spacecraft launched by the Air Force Space Test Program.
Unfortunately these spectrometers had no spatial resolution and
therefore it was not possible to observe evaporation in different
spatial locations.  Within two years more high resolution X-ray
spectrometers were flown on the NASA {\it Solar Maximum Mission}
spacecraft (e.g., Antonucci \& Dennis 1983) and the Japanese {\it
Hinotori} spacecraft (e.g. Tanaka 1986).  These missions confirmed the
evaporation signature but also lacked high spatial resolution.
Furthermore, the upflows were seen in only the highest temperature
lines, such as the \ion[Ca xix] line, and were not observed in cooler
lines due to poorer spectral resolution for the colder lines and lack
of coverage at low temperatures.  Thus, although the summed
characteristics of evaporation were well-observed in the early 1980s
for plasma at temperatures greater than about 10 MK, nothing was known
about the evaporation process at different loop footpoints and at
coronal temperatures near 1 MK.

The Bragg Crystal Spectrometer package flown in 1991 on the Japanese
{\it Yohkoh} spacecraft (Culhane et al. 1991) greatly improved
knowledge about the summed affects of evaporation on line profiles,
but did not offer any information concerning the spatial distribution
or locations of evaporation.  This changed with the launch of the
Coronal Diagnostics Spectrometer (CDS) (Harrison et al. 1995) on the ESA
{\it Solar \& Heliospheric Observatory}.  Several studies of
evaporation at specific locations were made with CDS at footpoint
locations in a line of Fe XIX at 592.23 \AA\ (i.e., Czaykowska et
al. 2001; Teriaca et al. 2003; Brosius \& Phillips 2004, and Milligan
et al. 2006a, 2006b).

The above data were interpreted with 1D hydrodynamic models (e.g.,
Nagai 1980; Cheng et al. 1983; Doschek et al. 1983; Nagai \& Emslie
1984; Emslie \& Nagai 1985) that eventually included loop tapering to
simulate constricting flux tubes near footpoints, and adaptive grid
meshes that allowed the chromosphere to move, although the transition
region could not be resolved due to the spatially rapid changing
temperature gradient that produces extremely small grid sizes.  Energy
and accelerated particles were introduced {\it ad hoc} anywhere
desired in the coronal part of the loop.  These early simulations
predicted that at flare onset the entire line profile of all
multi-million degree flare lines would be blueshifted.  This was not
observed in the summed flare data.  The CDS spectra obtained later
showed that these profiles were shifted, at least for \ion[Fe xix], at
flare footpoint regions.  However, the lack of a shift at the very
onset of flares was perplexing and contrary to model predictions
(e.g., Doschek et al. 1986; Doschek \& Warren 2005).

A solution of the problem of the stationary multi-million degree line
profiles at flare onset was first suggested by Hori et al. (1998).
These authors showed with a numerical simulation that if the
evaporation in a flux tube bundle took place sequentially in a number
of unresolved smaller flux tubes within the larger magnetic envelop,
then a stationary line profile could be be reproduced.  This
prediction was later tested with real flare data by Warren \& Doschek
(2005) and found to reproduce very well the observed BCS line profiles
from {\it Yohkoh}.

In summary, some of the Standard Flare Model predictions of the
appearance of chromospheric evaporation are consistent with
observations, but in general the total temperature coverage of
evaporation at specific flare locations has still remained small.
This has recently changed with the advent of the Extreme-ultraviolet
Imaging Spectrometer (EIS) flown on the Japanese {\it Hinode}
spacecraft.  This spectrometer observes well spectral lines formed at
1 MK up to about 20 MK.  It also observes a few lines of cooler
transition region lines from ions such as \ion[Fe viii] and \ion[He
ii].  Chromospheric evaporation at temperatures ranging from about
0.05 MK up to 16 MK was well observed in EIS spectra at loop
footpoints for a C1.1 class impulsive flare by Milligan \& Dennis
(2009).  They found that even at the footpoints the line profiles of
lines of multi-million degree \ion[Fe xxiii] and \ion[Fe xxiv] ions
contained a stationary component, which they concluded is a
disagreement with predictions of the Standard Model.  However, more
recent EIS spectra of a large number of flares have shown that in at
least some cases all the high temperature emission is moving upwards
at footpoints (Aoki \& Hara 2012; Brosius 2013).

With the end of the extended solar minimum, EIS is observing more and
more flares.  It requires good fortune for the EIS slit to be at a
significant flare location at a time when something significant might
be learned.  This has happened for an M1.8 flare that occurred on 9
March 2012.  The EIS slit was virtually on top of a significant energy
release site of the flare.  Evaporative upflows were observed at many
sites that we interpret as blended footpoint regions.  We have found
profiles for the multi-million degree \ion[Fe xxiii] line near 263
\AA\ that are completed shifted in wavelength towards the blue, almost
completely stationary in wavelength, and in some cases are a mixture
of shifted and unshifted components.  In this paper we discuss these
results and compare them with the results obtained by other authors
using EIS flare data.  Section 2 gives a brief summary of the EIS
instrument, Section 3 describes the flare and available observations,
results are given in Section 4, and a discussion and summary are given
in Section 5.

\section{THE EIS ON {\it HINODE} AND DATA REDUCTION}

EIS is described by Culhane et al. (2007) and Korendyke et al. (2006).
The instrument is a combination telescope and spectrometer.  The
telescope mirror and spectrometer grating are divided into two halves
and each is coated with different Mo/Si multi-layers.  Good
sensitivity is achieved over two extreme-ultraviolet narrow wavebands:
170-210 \AA\ and 250-290 \AA.  Light from the telescope mirror is
focused onto the entrance aperture of the spectrometer and passes
through to a grating which diffracts and focuses the light onto two
CCD detectors.  There are four entrance apertures that can be
selected: 1\arcsec\ slit, 2\arcsec\ slit, 40\arcsec\ slot, and
266\arcsec\ slot, all oriented in the north-south direction.

EIS can operate in raster mode, scanning from west to east, or in a
sit and stare mode, with differential rotation included.  At each
position within a raster it is possible to read-out the entire CCD
(full-CCD) and obtain a complete spectrum for each wavelength band or
alternatively one can select a small set of lines that fall in
narrower spectral windows.  The spatial resolution of EIS along the
slit is nominally 2\arcsec\ (1\arcsec\ per pixel) and the spectral
dispersion is 0.0223 \AA\ per pixel.  The instrumental full width at
half maximum (FWHM) is about 0.056 \AA.  The actual spatial resolution
is more like 3\arcsec\ due to the width of the point spread function.

The rastered flare data have been reduced using the standard EIS
software data reduction package, i.e., the data are corrected for
cosmic ray hits, warm pixels, detector bias, and dark current.  We
also corrected the wavelength positions for a variation of line
position over the {\it Hinode} orbit due to temperature variations in
the spectrometer and applied a correction for a slight tilt of the
slit on the CCDs, which shifts wavelengths as a function of the
north/south position of a pixel.

\section{OBSERVATIONS AND DATA REDUCTION}

The 9 March 2012 flare was a complex event that began in active region
11429 at N17, W12 near 03:24 UT.  It was a multiple event with a first
maximum at about 03:27 UT, where it reached class M1.8.  After a small
dip in intensity the X-ray flux continued to increase up to about
M6.8.  This increase occurred after the EIS slit had passed a portion
of the active region that had begun to strongly flare, producing the
additional intensity and an EIT wave.  The Perfect Storm scenario is
because the EIS slit was at or near a point of strong energy release
that produced the first flare maximum, as indicated by the 193 \AA\
filter channel on the Atmospheric Imaging Assembly telescopes flown on
the {\it Solar Dynamics Observatory} ({\it SDO}).  This channel showed
a strong CCD saturation (bleeding in the north/south direction and
highly localized in the east/west direction) very close to the
location of the EIS slit.  Unfortunately, after this energy pulse many
of the AIA images are badly saturated for some time as the flare
evolves.  Neverthless, the first images are revealing regarding the
locations of the strongest energy inputs.  The location of the energy
input observed by EIS is shown in Figure~\ref{f1.eps} along with GOES
X-ray data for the flare.  The vertical lines show the location of the
EIS slit when a significant increase in flare intensity occurred.  EIS
scans from west to east (right to left in the \ion[Fe xiii] line image
shown).

The flare data were obtained by rastering the 2\arcsec\ slit from west
to east across an active region in 2\arcsec\ steps with a 30s exposure
time at each step.  The raster covered 120\arcsec\ in the west-east
direction (60 raster positions) and the slit length (north-south
direction) was 160\arcsec.  Monochromatic images of solar regions were
obtained by rastering the slit, integrating the spectra over
wavelength, and then stacking the spectra side-by-side.

The particular active region raster study chosen for the 9 March
active region was a full-CCD study, and therefore every spectral line
observable by EIS was recorded.  This not only allows examination of
lines formed over a broad range of temperatures, but it also helps
with saturation problems.  For example, at the first pulse peak, the
192 \AA\ \ion[Fe xxiv] line is saturated, but the substantially weaker
263 \AA\ \ion[Fe xxiii] is not saturated.  Thus, saturation effects
are not as severe when weak lines from other ions or from the same ion
can be used in place of saturated stronger ones.

A number of spectral lines were chosen for inspection and for
analyzing evaporative upflows.  They are listed in Table 1 with their
wavelengths and temperatures of formation assuming ionization
equilibrium.  The temperatures are from the ion balance given in
CHIANTI (Dere et al. 2009, A\&A, 498, 915).

Inspection of the line profiles shows that the lines of multi-million
degree ions such as \ion[Fe xxiii] have significant blueshifted
components in some parts of the active region.  This is where
chromospheric evaporation is occurring along the same line-of-sight as
plasma that has already reached a stationary state, i.e., no net
Doppler speed.  Since multiple upflowing sources are probably present
along the same-line-of-sight, we simply divide the line profile into
two Doppler speed intervals (see top left panel of
Figure~\ref{f6.eps}), rather than attempt a multiple fit with Gaussian
line profiles.  We then make images out of the centroid component and
the two Doppler shifted asymmetric components.  We don't know how many
Gaussians to use.  However, the lines from cooler coronal ions such as
\ion[Fe x] - \ion[Fe xvi] do not show the broad blueshifted components
and can therefore be fit with single Gaussians.  Thus for each of
these lines we obtain a total line intensity, a Doppler speed (from
the centroid wavelength), and a line width (full width at half
maximum, FWHM).

The FWHM is assumed to be due to instrumental broadening (known),
thermal Doppler broadening, and excess broadening due to turbulence or
just random non-thermal motions.  The non-thermal broadening $V$ is
measured assuming Gaussian distributions for all components, i.e.,
\begin{equation}
FWHM = 1.665\times10^3~\frac{\lambda}{c}~\sqrt{\frac{2kT}{M} +
V^2 + W^2_I}~~~,
\end{equation}
where $\lambda$ is the wavelength (in \AA\ in this paper), $c$ is the
speed of light, $k$ is the Boltzmann constant, $T$ is the electron
temperature, and $M$ is the ion mass.  The instrumental width $W_I$
(also assumed Gaussian) is approximately 68 m\AA\ for the 2\arcsec\
slit, using the IDL routine eis\_slit\_width for the approximate
location of the flare spectrum on the detector.  Furthermore, if
ionization equilibrium is assumed, there is the implicit assumption
that the electron and ion temperatures are equal.  This assumption is
likely valid in the flare where the EIS observations are made, because
electron densities in flares in the corona are known to be quite high
(on the order of several times 10$^9$ to several times 10$^{11}$
cm$^{-3}$.  At these densities equilibration times between electrons
and ions are very short, and ionization and recombination processes
are very rapid for ions such as \ion[Fe xii].

Evaporation upflows or downflowing plasma speeds are determined from
the wavelengths of the lines.  Unfortunately EIS does not contain an
absolute wavelength scale.  We use rest wavelengths determined from
off-limb observations by Warren et al. (2011) and assume that the
uppermost and/or bottom part of the raster is essentially not moving
along our line-of-sight.  This region is well-outside of the flare
region.

Electron densities to be discussed below were determined from the
\ion[Fe xiii] line ratio, 203.83/202.04 \AA\ discussed by Young et
al. (2009) and an \ion[Fe xiv] ratio, 264.79/274.20 \AA.  Using data
in CHIANTI, the \ion[Fe xiii] densities differ from densities obtained
from an \ion[Fe xii] line ratio used frequently in EIS papers (see
Young et al. 2009) by about a factor of 2.5.  This discrepancy is
heading towards a resolution with revised atomic data for \ion[Fe xii]
(Del Zanna 2012) and also for \ion[Fe xiii].  These data are not yet
in CHIANTI but will appear in the next CHIANTI release.  According to
Young et al. (2012), with the old \ion[Fe xiii] and \ion[Fe xii]
atomic data a log10 density of 8.6 determined from the \ion[Fe xiii]
ratio would be a log10 density of 8.9 using the \ion[Fe xii] data.
With the new atomic data, these ratios give log10 densities of 8.2
(\ion[Fe xiii]) and 8.44 (\ion[Fe xii]) and the error bars now
overlap.

\section{RESULTS}

\subsection{Sites of Evaporating Plasma - Overview}

Figure~\ref{f2.eps} shows maps of the Doppler speed calculated from
the centroid wavelengths of selected coronal spectral lines.  In these
images the rasters are put side by side, and are not separated by
2\arcsec\ as in Figure~\ref{f1.eps}. Figure~\ref{f1.eps} is a correct
image representation for comparison with AIA images, but
Figure~\ref{f2.eps} shows the correct pixel numbers in the east-west
direction and is therefore easier to relate to the EIS parameters.
Another point to note is that the images are represented as they
appear on the long wavelength EIS CCD.  The short wavelength images
have been shifted in the north-south direction by 17 pixels to be
spatially aligned with the long wavelength images.  The images
concentrate on the flare region and are bounded by the yellow box in
Figure~\ref{f1.eps}.  In the figure evaporating sites appear dark
indicating negative Doppler speed.  Note that all the lines show
evaporation.  For display the \ion[Fe xxiii] Doppler map is made by
subtracting the two blue-shifted component images from the centroid
image after multiplying the blue-shifted intensities by 2.  The
sources of footpoints are not as clear as for the smaller event
discussed by Milligan \& Dennis (2009).

Images of the flare in selected spectral lines are shown in
Figure~\ref{f3.eps} through Figure~\ref{f6.eps}, shown with rasters
spaced by 2\arcsec\ as in Figure~\ref{f1.eps}.  Transition region line
images are shown in Figure~\ref{f3.eps}, coronal line images are shown
in Figure~\ref{f4.eps} and in the top panels of Figure~\ref{f5.eps}.
The bottom panels of Figure~\ref{f5.eps} and Figure~\ref{f6.eps} show
multi-million degree flare line images.  In Figure~\ref{f6.eps} the
images represent the emission bounded by the velocity intervals shown
in the \ion[Fe xxiii] line profile panel.  In Figure~\ref{f5.eps} and
Figure~\ref{f6.eps} the mottling in the bright regions for the \ion[Fe
xv] and \ion[Fe xxiv] images is probably due to some saturated pixels.

Note that many of these figures show a small feature between 0 and 10
in X-pixels and 15 an 20 in Y-pixels.  This feature is brightest in
transition region ions and is believed to be a fairly cold region
although there is apparently a coronal component.  Therefore, this
emission in the bottom right panel of Figure~\ref{f6.eps} may not be
due to \ion[Fe xxiii] but may instead be due to weak unidentified
lines in the blue wing of the \ion[Fe xxiii] line.  This is
particularly true for the 260 - 470 km s$^{-1}$ image.  Weak
unidentified lines far from the centroid of an emission line of
interest can sometimes lead to confusion if the line of interest has a
large blueshifted or redshifted component.  See also Del Zanna et
al. (2011) for a discussion of other weak lines that blend the short
wavelength region of the \ion[Fe xxiii] line.

The short vertical lines in Figure~\ref{f3.eps} through
Figure~\ref{f5.eps} indicate X-pixel locations (twice the actual pixel
number) for which we show the evaporative speeds in detail.
Figure~\ref{f7.eps} shows the Doppler speed plotted for four selected
X-pixels as a function of Y-pixel for the entire raster in the
north-south direction. The X-pixel = 43 was observed before the flare
erupted in this region and shows no net strong flows.  (The X-pixel =
43 is not displayed in Figure~\ref{f3.eps} through
Figure~\ref{f5.eps}.)  Similarly, no strong flows are seen for X-pixel
= 27, but evaporative upflows between 20 and 60 km s$^{-1}$ are seen
for X-pixels 12 and 19 (pixels 24 and 38 in Figure~\ref{f3.eps}
through Figure~\ref{f5.eps}) for lines of the ions indicated.

The variations of the upflow are shown in more detail in
Figure~\ref{f8.eps} and Figure~\ref{f9.eps}.  In these figures the
upflows are shown for eight consecutive X-pixels for four spectral
lines, formed in typical active region loops.  Note that the flows
change significantly over a few arcsecond distances in both X and Y.

The above results are concerned with coronal lines and the transition
region ion, \ion[Fe viii].  The upflows seen in \ion[Fe xxiii] and
\ion[Fe xxiv] are much greater.  Sample \ion[Fe xxiii] line profiles
are shown in Figure~\ref{f10.eps}.  The inset image shows the
locations of the profiles.  \ion[Fe xxiii] images with 2\arcsec\
intervals between rasters are shown in Figure~\ref{f5.eps} and
Figure~\ref{f6.eps}.  These are the images to compare with AIA images.
The numbers in parentheses in Figure~\ref{f10.eps} are the X-pixel and
Y-pixel numbers, respectively.  The intensity of the image at (24,25)
has been divided by two to fit within the plot.  The peak intensities
of the lines near X-pixel 24 show little or no blueshift and we assume
that they are at or close to the rest wavelength.  Most of the upflows
are seen east of this position.

The eastward progression of the upflows can be seen in
Figure~\ref{f11.eps}.  The Figure shows a selected path in the image
and the upflow speeds obtained from the line centroids along this
path, with position 0 near X-pixel = 12 and position 10 near X-pixel
position 27.  The upflows here are not the blue-shifted wing of a
profile, but are the shifts of the peak intensities of the profiles.
Note that the coronal upflows are much less than the multi-million
degree upflows.  Transition region lines also show small upflows.

\subsection{Sites of Evaporating Plasma - a Quantitative Example}

We provide some quantitative results for hydrodynamic modeling in
Figure~\ref{f12.eps}, Figure~\ref{f13.eps}, and Figure~\ref{f14.eps}
for one spectrum obtained at X-pixel = 12.  In these figures we give
the Doppler speeds, intensities, and full widths at half maxima
(FWHMs) for the lines in Table 1 for \ion[Fe viii], \ion[Fe x],
\ion[Fe xii], \ion[Fe xiii], \ion[Fe xiv], \ion[Fe xv], \ion[Fe xvi],
and \ion[Fe xxiii].  The inset vertical line in Figure~\ref{f12.eps}
shows the range of Y-pixels included in the plots.  Also, in
Figure~\ref{f13.eps} we give densities obtained from density sensitive
intensity ratios of lines of \ion[Fe xiii] and \ion[Fe xiv].  The
\ion[Fe xxiii] lines in Figure~\ref{f14.eps} are mostly Gaussian in
shape but shifted from their rest positions.  However, in some cases
there is a slight blueshifted wing.  Therefore we have fit the \ion[Fe
xxiii] lines with both single Gaussians and with two Gaussians.  All
the data in Figure~\ref{f14.eps} are for the main Gaussian component.
The image in Figure~\ref{f14.eps} shows the total \ion[Fe xxiii]
emission as well as the emission at wavelengths shortward of the rest
wavelength, 263.76 \AA.  In the top left panel we show the results for
both one and two Gaussian fits.  They look quite similar.  The one
Gaussian fit gives slightly larger speeds as expected, and the two
Gaussian fit shows more detail, also as expected.  However, this
detail may not be real.  Finally, we have given FWHMs instead of
non-thermal speeds in case a modeler does not wish to assume
ionization equilibrium.  Non-thermal speeds can be calculated using
equation (1) and whatever temperature is assumed for the formation of
the spectral line.

\subsection{The First Flare Peak}

We have noted that there was a large energy input near 03:26 UT that
appears as a spatially-small, highly localized brightening in the 335
AIA channel.  This occurred about 30s before the EIS slit reached this
position, and therefore it is interesting to explore the dynamics of
spectral lines near this location, as well as temperatures for the
multi-million degree flare plasma.  The coordinates of maximum \ion[Fe
xxiii] intensity are X = 27, Y = 75, at 03:26:27 UT.

Figure~\ref{f15.eps} shows \ion[Fe xxiii] line intensities for X and Y
pixels near (27,75) and line profiles of \ion[Fe xxiii], \ion[Fe
xiii], and \ion[Fe xiv].  Note that the \ion[Fe xxiii] emission is
highly localized but that there is only a hint of a blue-wing on the
line profile.  The position is not a site of high velocity evaporation
in the \ion[Fe xxiii] line.  Perhaps much more interesting are the
red-wings that appear on the \ion[Fe xiii] and \ion[Fe xiv] line
profiles.  Since these lines were fit by a single Gaussian, the effect
on the fit is to shift the centroids to the red, showing a downflow.
This is seen in Figure~\ref{f16.eps} where the Doppler speeds are
shown for four selected X-pixels around (27,75).  It is remarkable
that the Doppler speeds of the two coronal lines track each other so
well, even for small variations that might ordinarily be found to be
noise.  The two lines are not even from the same detector.  The
downflow is quite localized in position, and is striking for X-pixel
28.  The EIS slit at (27-28,75) is very close in space and time to a
sunquake (Kosovichev 2012).  In the next section we speculate on
possible connections of the coronal downflows observed in this region
and the sunquake.

At location (27,75) five high temperature flare lines can be used to
construct the loci plots in Figure~\ref{f17.eps} that give information
on the differential emission measure of the source (e.g., see Warren
\& Brooks 2009 for and explanation of loci plots).  The figure shows
loci plots for lines of \ion[Fe xvii], \ion[Fe xxii], \ion[Fe xxiii],
and \ion[Fe xxiv].  Again remarkably the highest temperature lines
cross at virtually the same location, which indicates an isothermal
source at about $14\times10^6$ K.  The \ion[Fe xvii] line is formed at
much lower temperatures and is formed outside the isothermal source.

\section{DISCUSSION}

The results in Section 4 show that the 9 March flare is quite complex
in both space and time.  EIS has observed upflow regions in
unprecedented detail in both spatial and temperature coverage.  In
attempting to make sense of these observations it is important to keep
in mind that the time resolution of EIS is low compared to the
impulsive time scales of the beginning of this flare.  The EIS
exposure times are 30s, so that for example the eight X-pixels for
which data are shown in Figure~\ref{f8.eps} and Figure~\ref{f9.eps}
cover four minutes of time in X.  However, all the data along Y are
simultaneous.

An attempt can be made to understand the temporal/spatial evolution of
the event by examining AIA filter data.  However, we find that at
flare onset some of the data is overexposed, and it is difficult to
separate high temperature plasma from ions such as \ion[Fe xxiv] from
the cooler coronal emission that occurs in the same filter bandpass.
Nevertheless, inspection of post-flare loops very late in the decay
phase indicate that loops observed by EIS at flare onset were highly
sheared relative to the neutral line (HMI data) and were basically in
an east-west direction.

\begin{figure}[t!]
\plotone{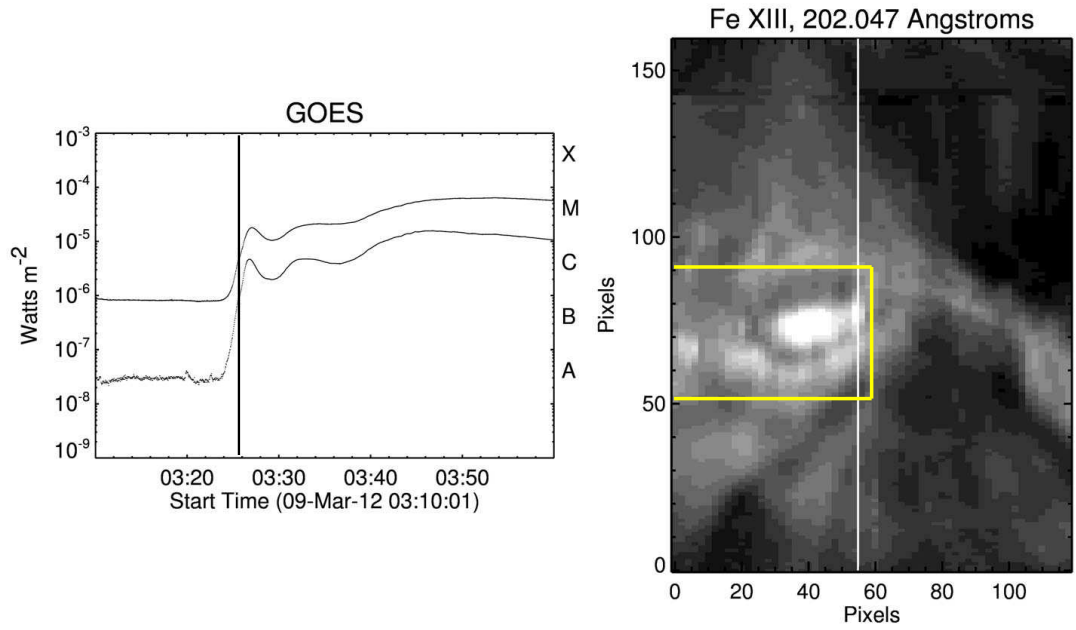}
\caption{\label{f1.eps} The location of the EIS slit (vertical lines)
  in an EIS image of the flare and relative to X-ray emission recorded
  by GOES.  EIS scans from west to east (right to left in the image).}
\end{figure}

From inspection of all the figures in the previous section, we can
conclude that:

1. The flare began before EIS observations began (see
Figure~\ref{f1.eps}).  GOES data show that EIS began observing during
the initial impulsive phase which was well-underway when EIS observing
started.

\begin{figure}[t!]
\plotone{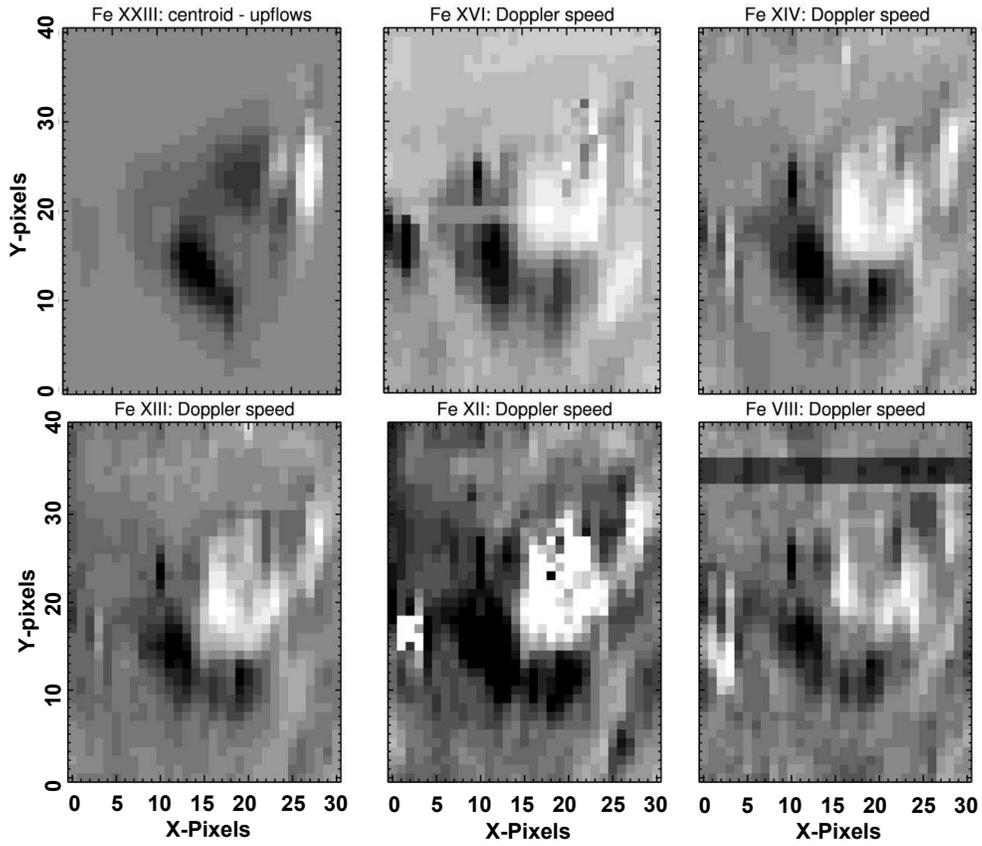}
\caption{\label{f2.eps} Centroid maps of selected spectral lines.  The
maps are bounded by the yellow box in Figure~\ref{f1.eps}.  Dark
indicates an upflow or evaporating plasma.}
\end{figure}

\begin{figure}[t!]
\plotone{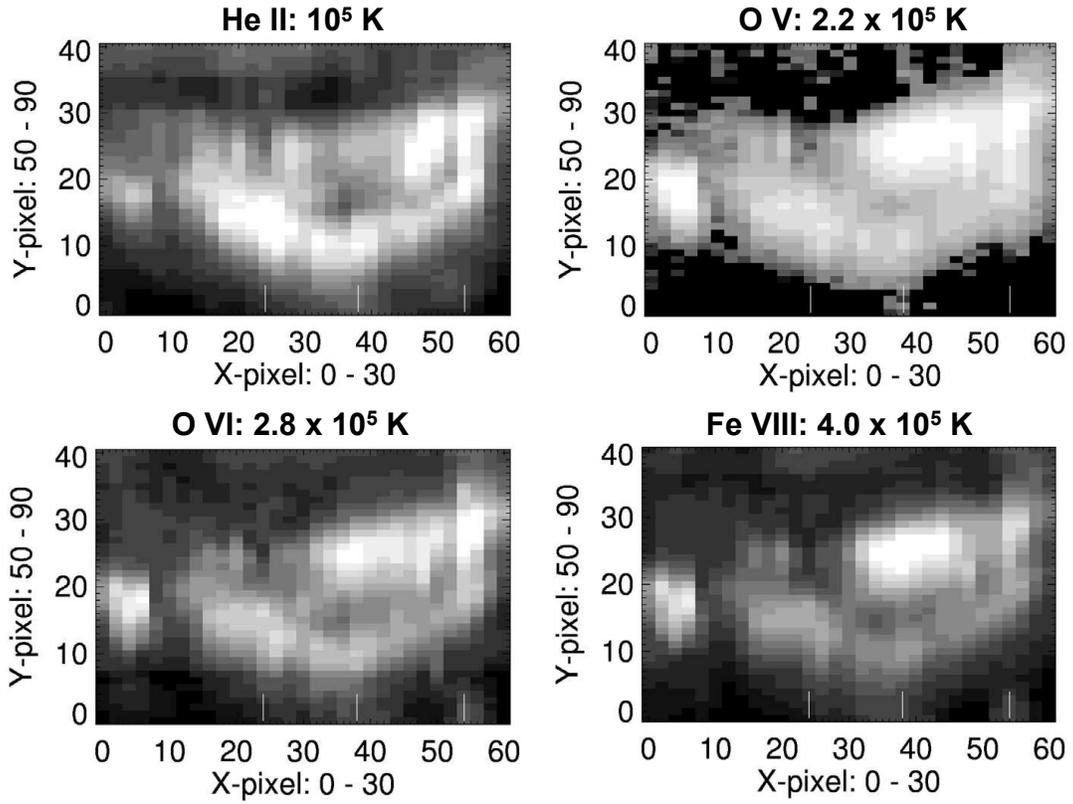}
\caption{\label{f3.eps} Transition region line images of the flare
  region bounded by the yellow box in Figure~\ref{f1.eps}.  The short
  vertical lines refer to X-pixels for which detailed evaporation
  speeds are shown in Figure~\ref{f7.eps}.}
\end{figure}

\begin{figure}[t!]
\plotone{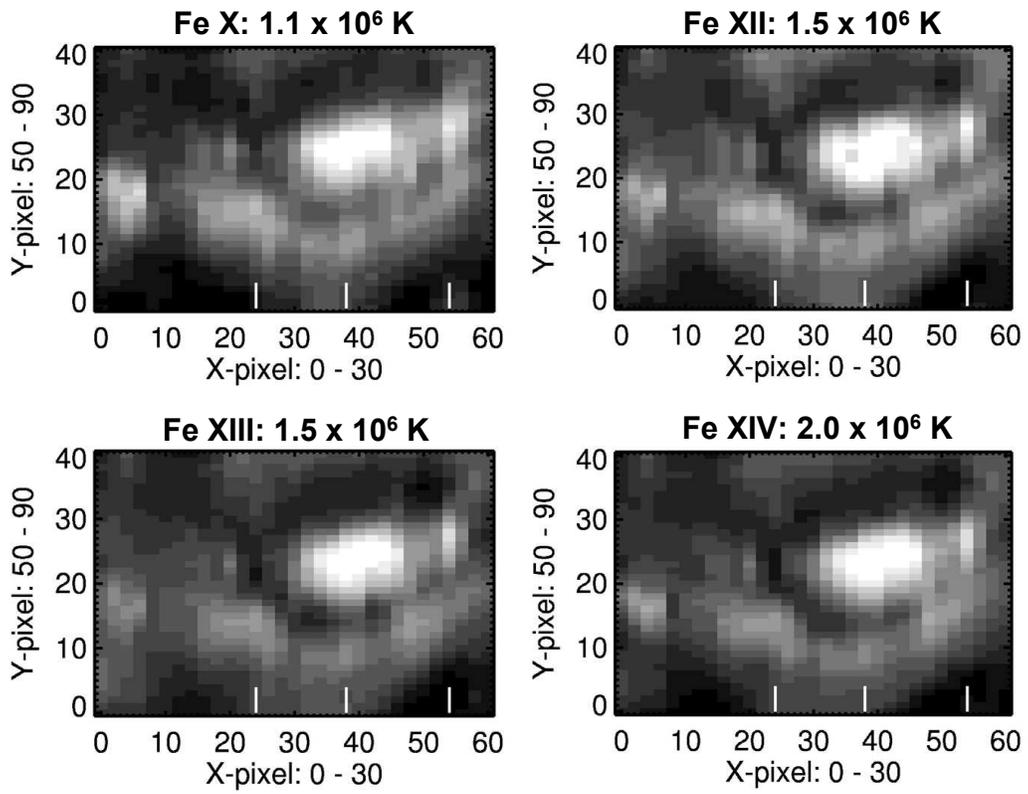}
\caption{\label{f4.eps} Coronal line images of the flare similar to
Figure~\ref{f3.eps}.}
\end{figure}

\begin{figure}[t!]
\plotone{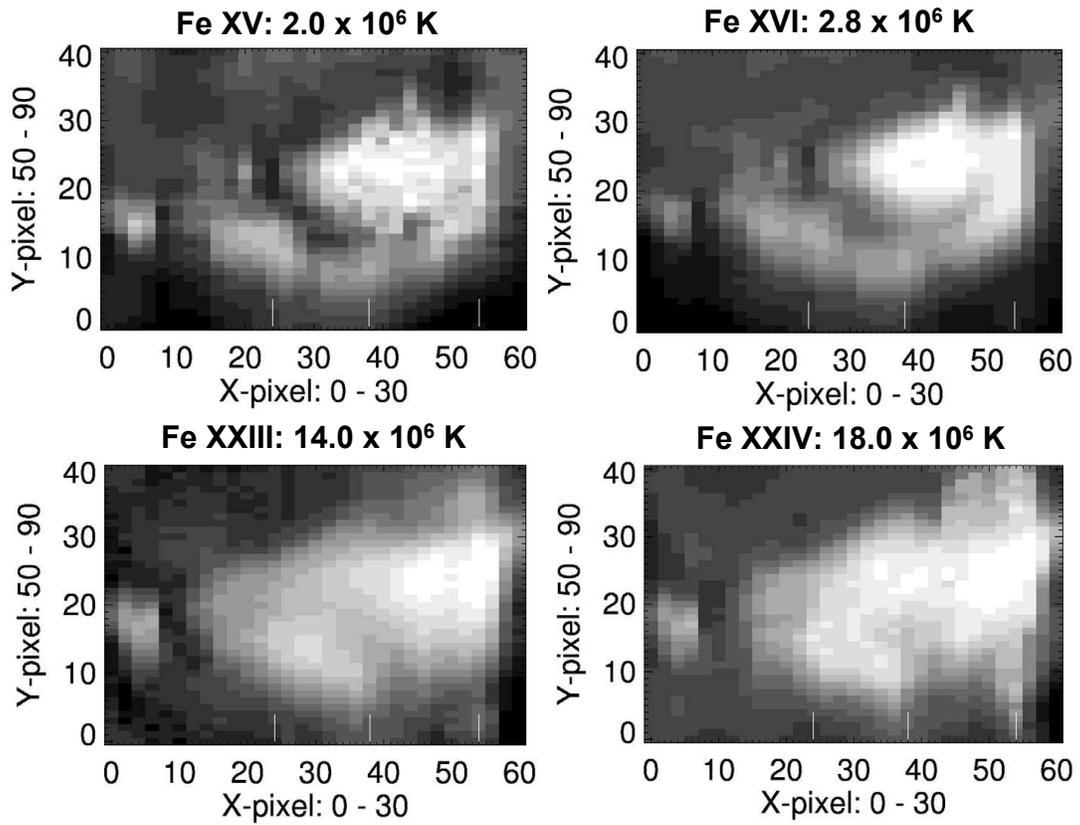}
\caption{\label{f5.eps} Coronal line and multi-million degree flare
line images of the flare similar to Figure~\ref{f3.eps}.}
\end{figure}

\begin{figure}[t!]
\plotone{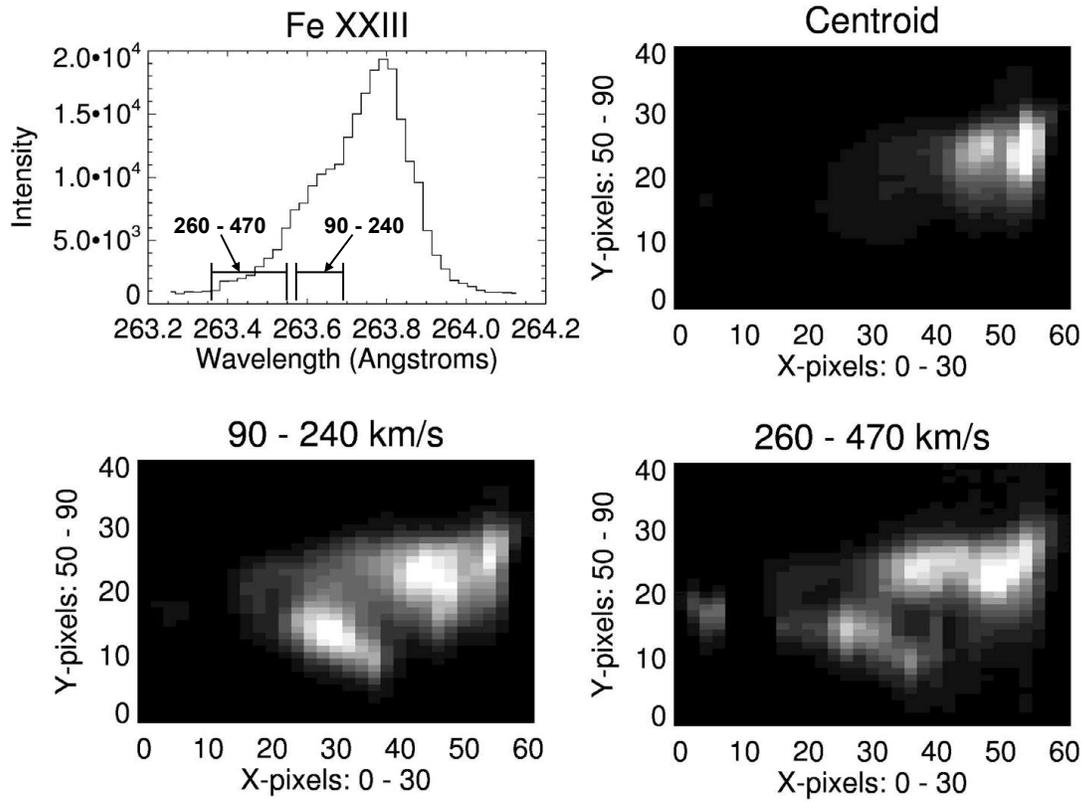}
\caption{\label{f6.eps} \ion[Fe xxiii] images of the flare in the
  velocity intervals 90 - 240 and 260 - 470 km s$^{-1}$ defined by the
  \ion[Fe xxiii] line profile in the top left panel.}
\end{figure}

\begin{figure}[t!]
\plotone{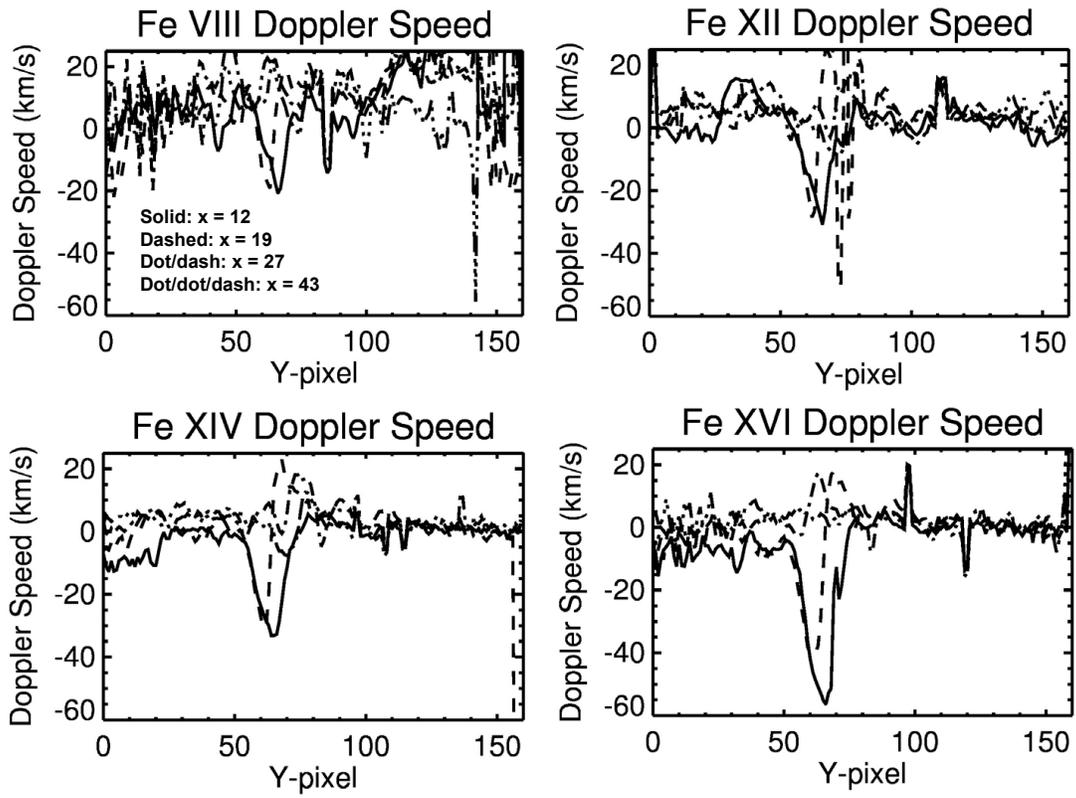}
\caption{\label{f7.eps} Evaporative speeds for lines of the ions
  indicated for 4 selected X-pixels.  Negative speeds are upflows
  towards the observer.}
\end{figure}

\begin{figure}[t!]
\plotone{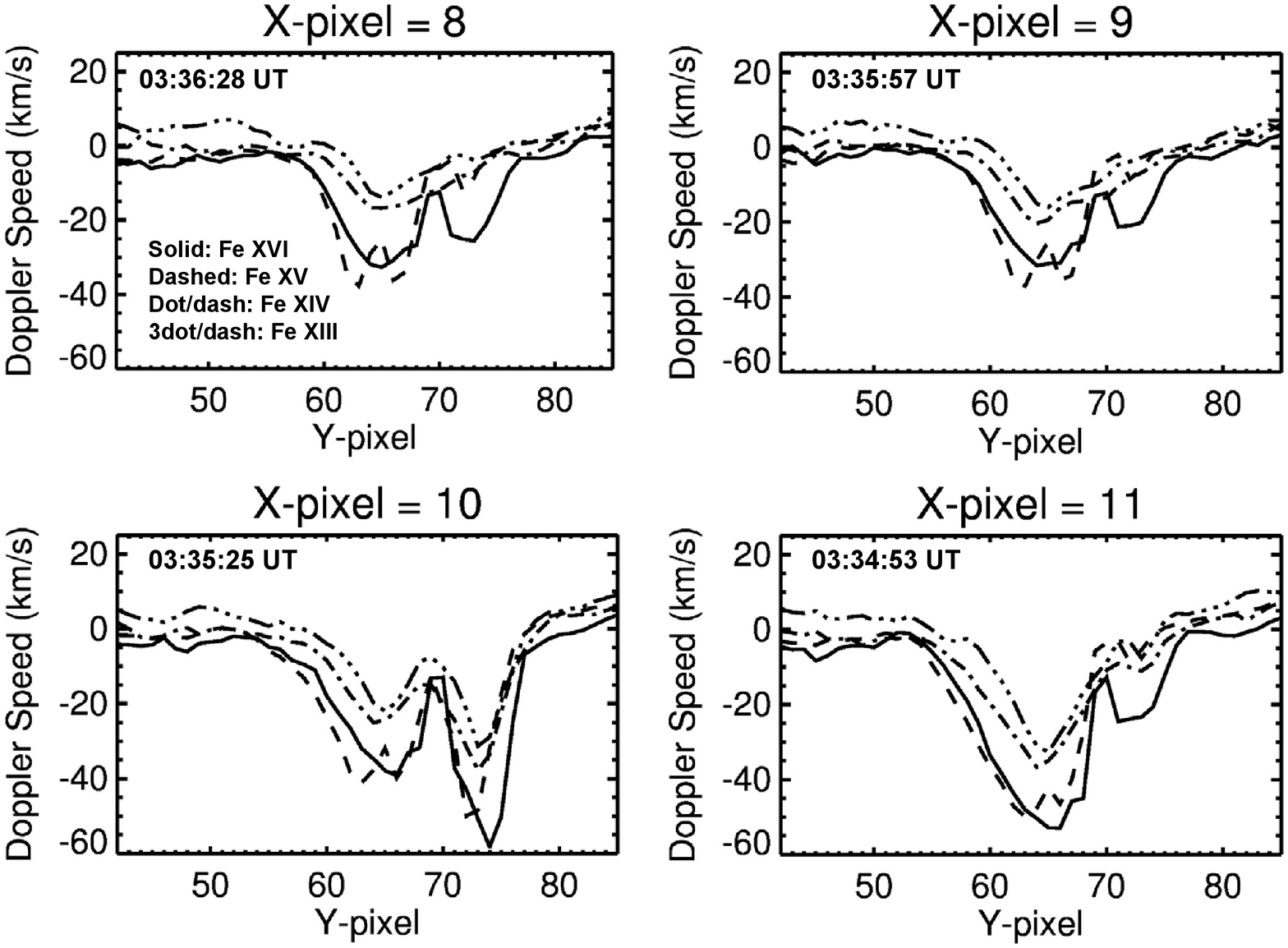}
\caption{\label{f8.eps} Details of upflows for selected X-pixels in
  the lines of \ion[Fe xiii], \ion[Fe xiv], \ion[Fe xv] and \ion[Fe xvi].}
\end{figure}

\begin{figure}[t!]
\plotone{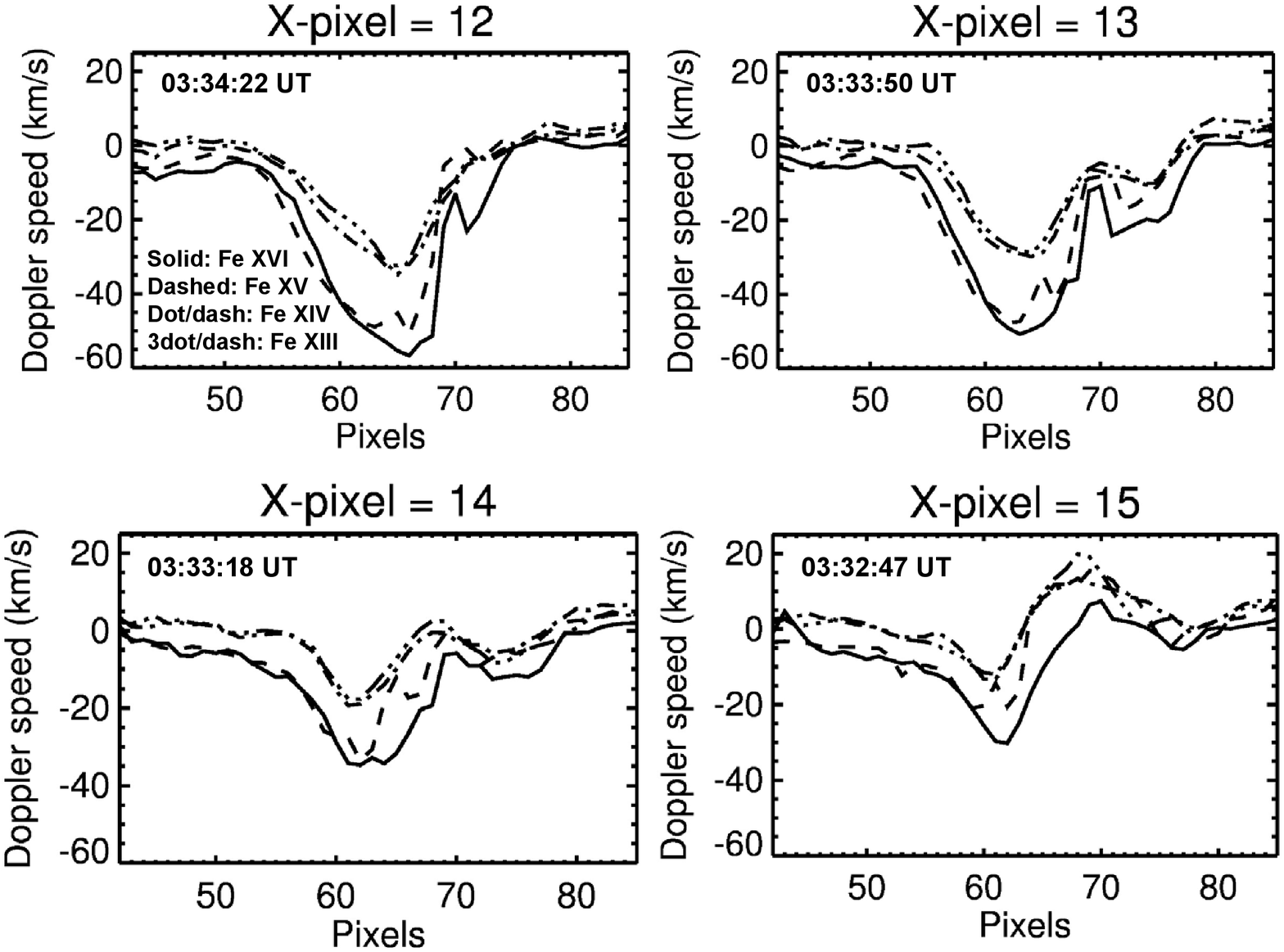}
\caption{\label{f9.eps} Details of upflows for selected X-pixels in
  the lines of \ion[Fe xiii], \ion[Fe xiv], \ion[Fe xv] and \ion[Fe xvi].}
\end{figure}

\begin{figure}[t!]
\plotone{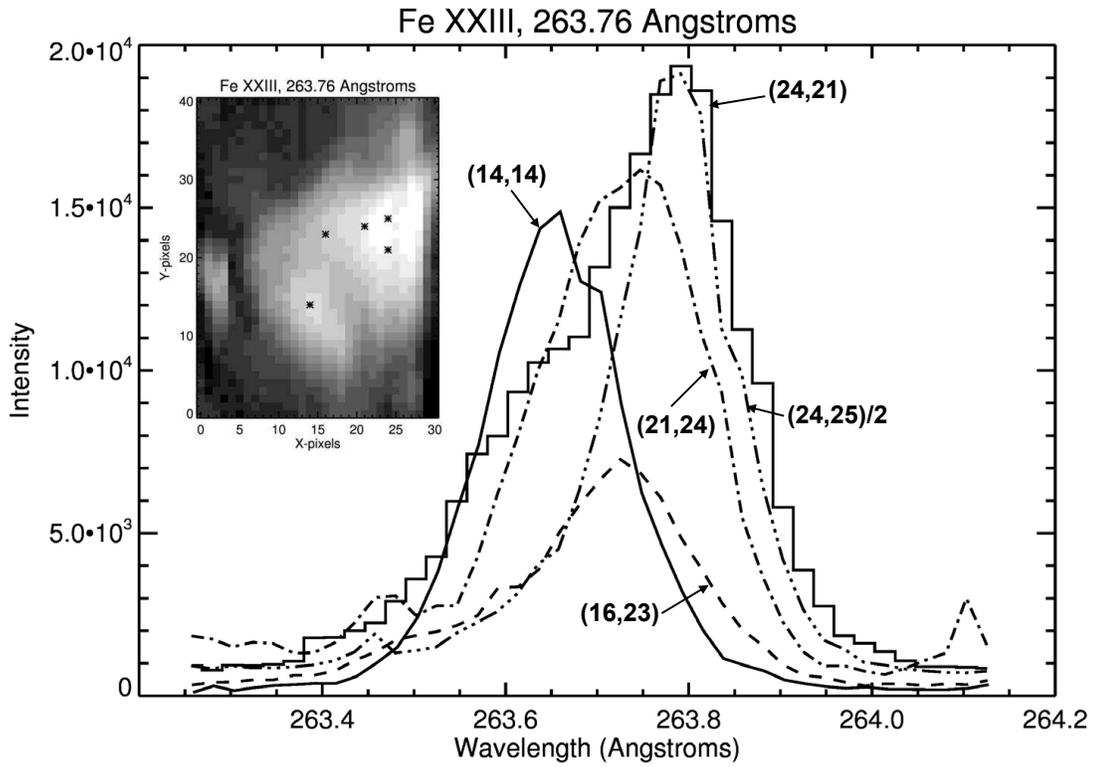}
\caption{\label{f10.eps} Line profiles of the \ion[Fe xxiii] line at
  selected X and Y pixels.  The numbers in parentheses are the X and Y
  pixels, respectively.  The intensity of the (24,25) profile has been
  divided by 2 to fit onto the plot. The inset image shows the pixel
  positions (see also the 2\arcsec\ raster position images in
  Figure~\ref{f5.eps} and Figure~\ref{f6.eps} for comparison with AIA
  images.)}
\end{figure}

\begin{figure}[t!]
\plotone{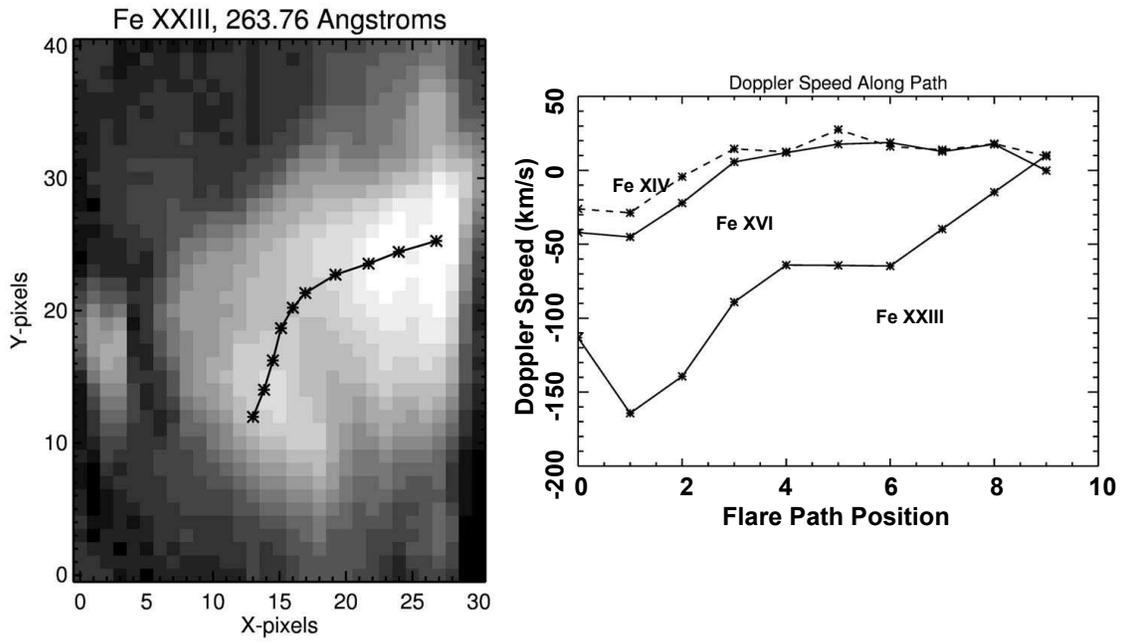}
\caption{\label{f11.eps} The Doppler speeds of the lines of the three
  ions shown as a function of position along the path.  Position 0 is
  near X-pixel = 12.}
\end{figure}

2. Figure~\ref{f2.eps} shows that extensive outflow and downflow
regions were observed between approximately X = 0 (03:40:42 UT) to X =
30 (03:24:52 UT) for coronal lines and the transition region line of
\ion[Fe viii].  \ion[Fe xxiii] showed outflows almost over the entire
raster between X = 0 and X = 30, due to the presence of blueshifted
asymmetries on line profiles as well as blueshifts of the peak
emission of the entire line profile.

\begin{figure}[t!]
\plotone{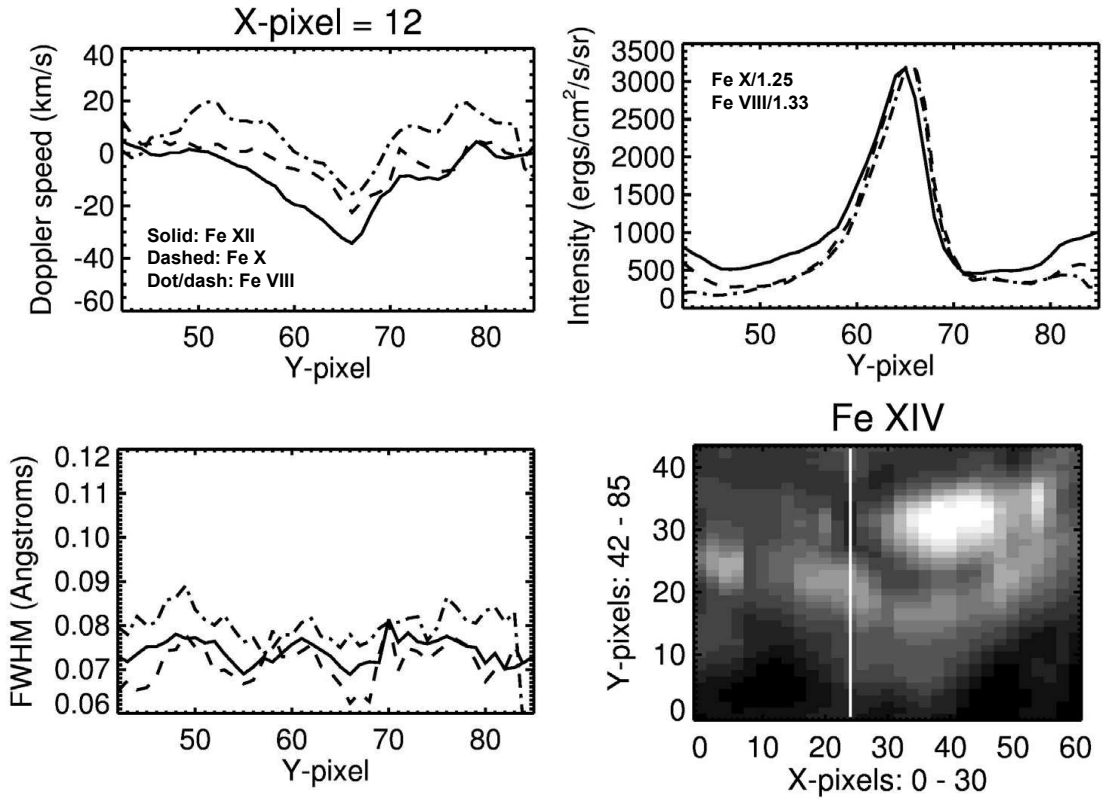}
\caption{\label{f12.eps} Doppler speed, intensity, and FWHM for
the upflow region at X-pixel = 12.  The intensities of the \ion[Fe
viii] and \ion[Fe x] lines are normalized to the intensity of the
\ion[Fe xii] line.  The vertical line in the \ion[Fe xiv] image shows
the range of Y-pixels included in the plots.}
\end{figure}

\begin{figure}[t!]
\plotone{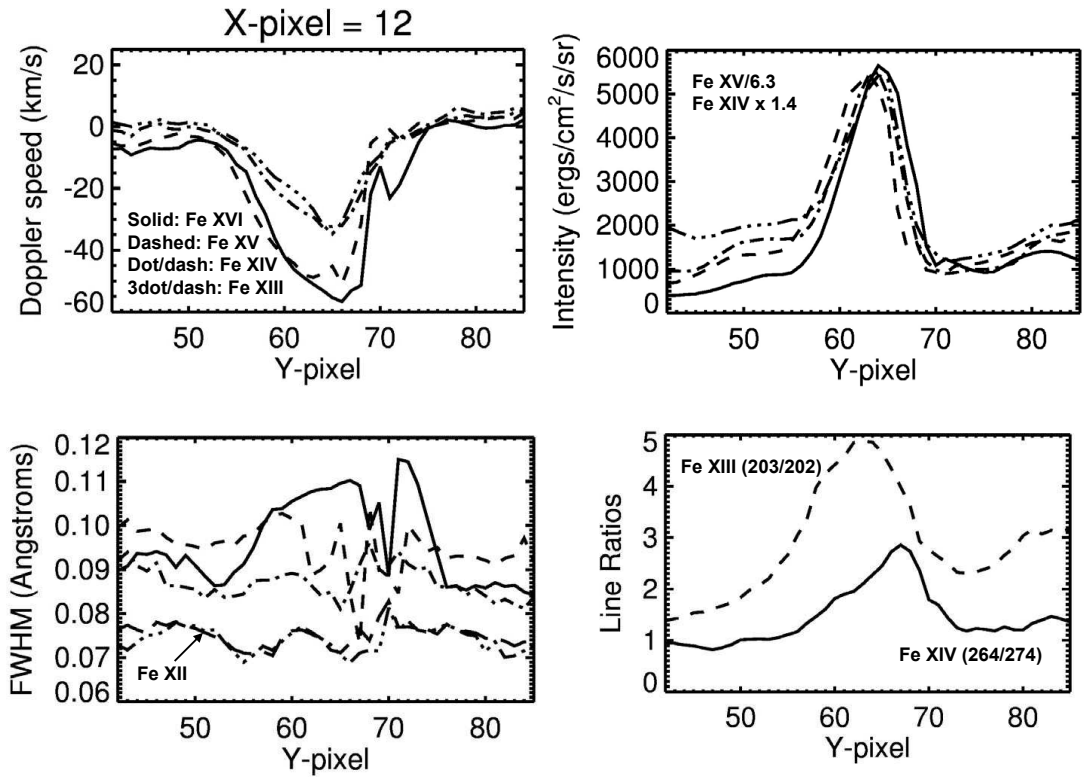}
\caption{\label{f13.eps} Similar to Figure~\ref{f12.eps} for
  lines of \ion[Fe xiii] - \ion[Fe xvi].  The normalized intensities
  are indicated.  The lower right plot gives density sensitive line
  ratios for \ion[Fe xiii] and \ion[Fe xiv] lines.}
\end{figure}

\begin{figure}[t!]
\plotone{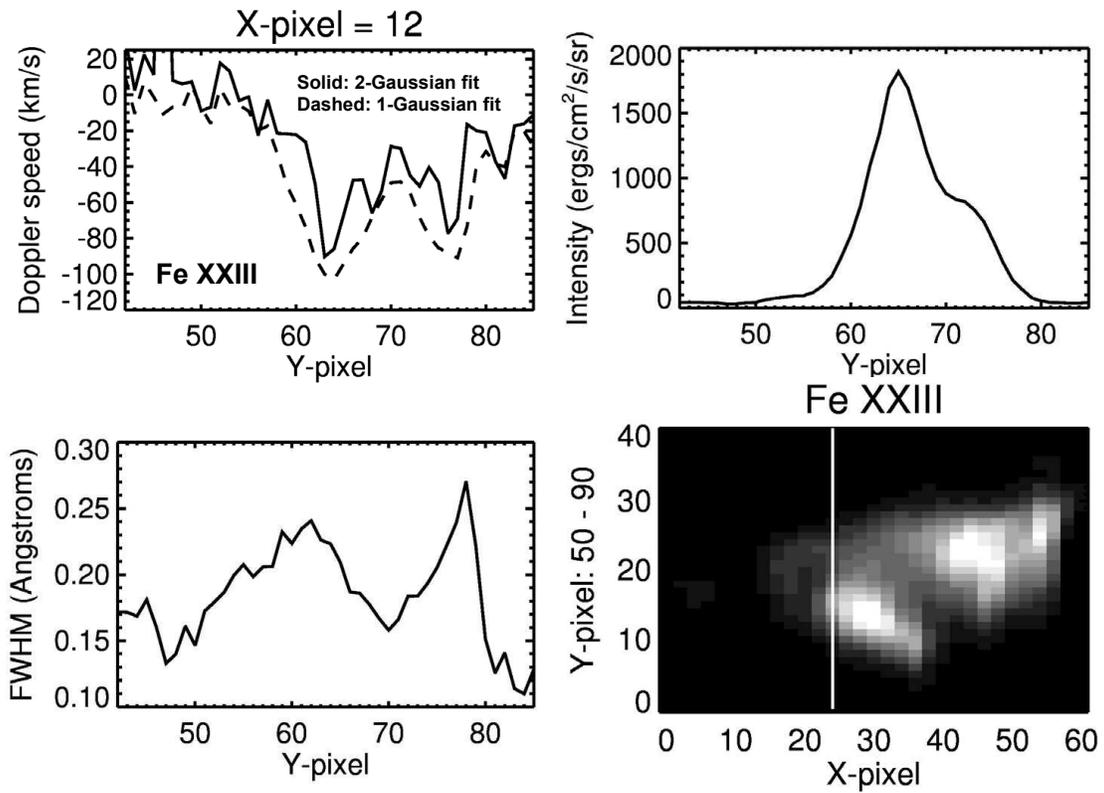}
\caption{\label{f14.eps} Similar to Figure~\ref{f12.eps} for
  the \ion[Fe xxiii] line.}
\end{figure}

\begin{figure}[t!]
\plotone{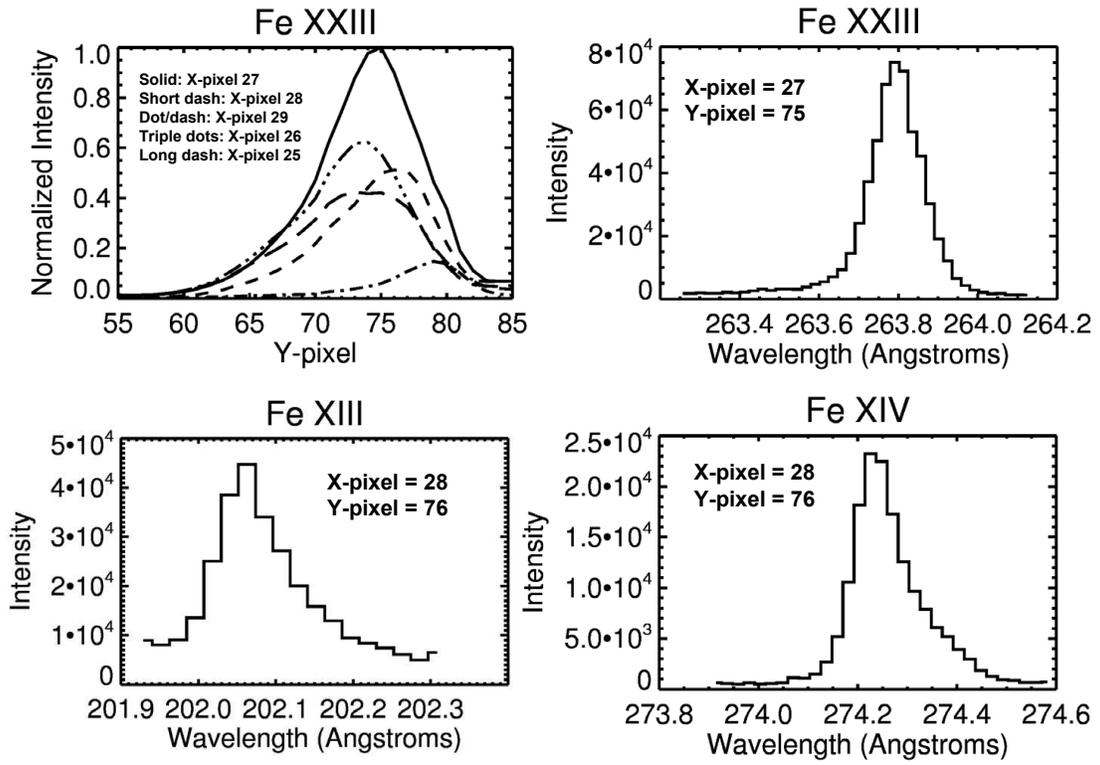}
\caption{\label{f15.eps} \ion[Fe xxiii] line intensities and selected
  line profiles for the indicated lines around position (27,75), a
  location close to, or at, a significant flare energy input.}
\end{figure}

\begin{figure}[t!]
\plotone{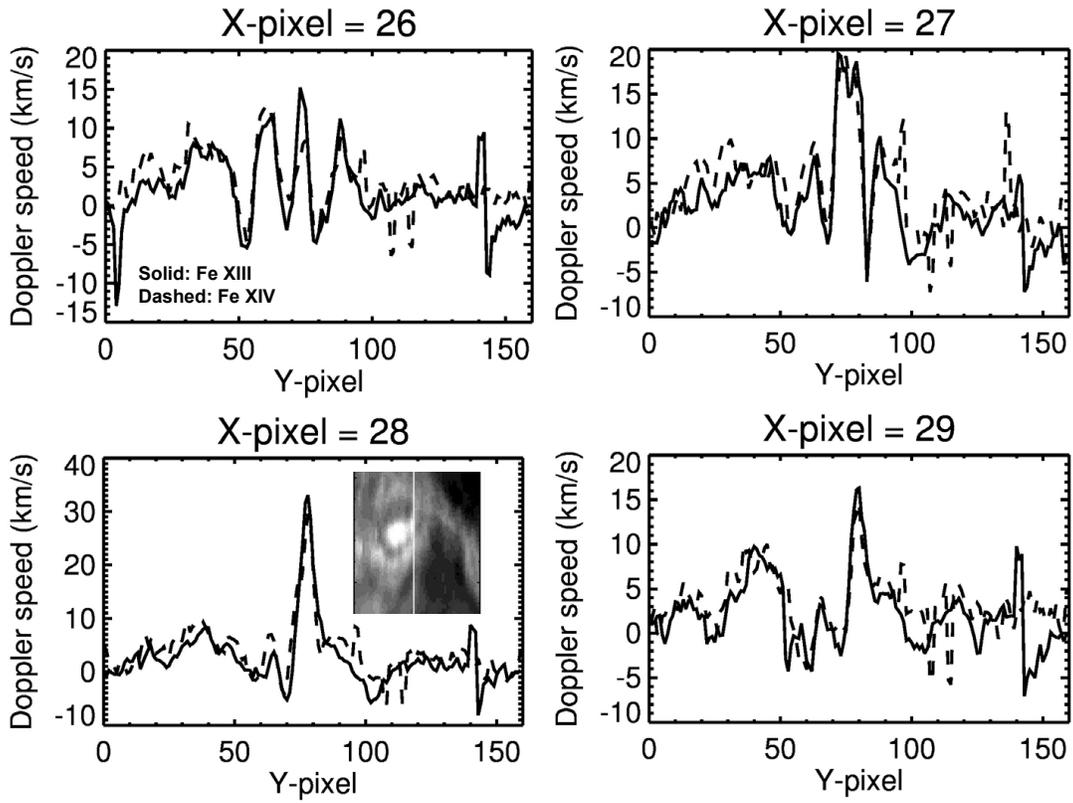}
\caption{\label{f16.eps} Doppler speeds for the indicated lines at
  four selected X-pixels for all Y-pixels associated with them.}
\end{figure}

\begin{figure}[t!]
\plotone{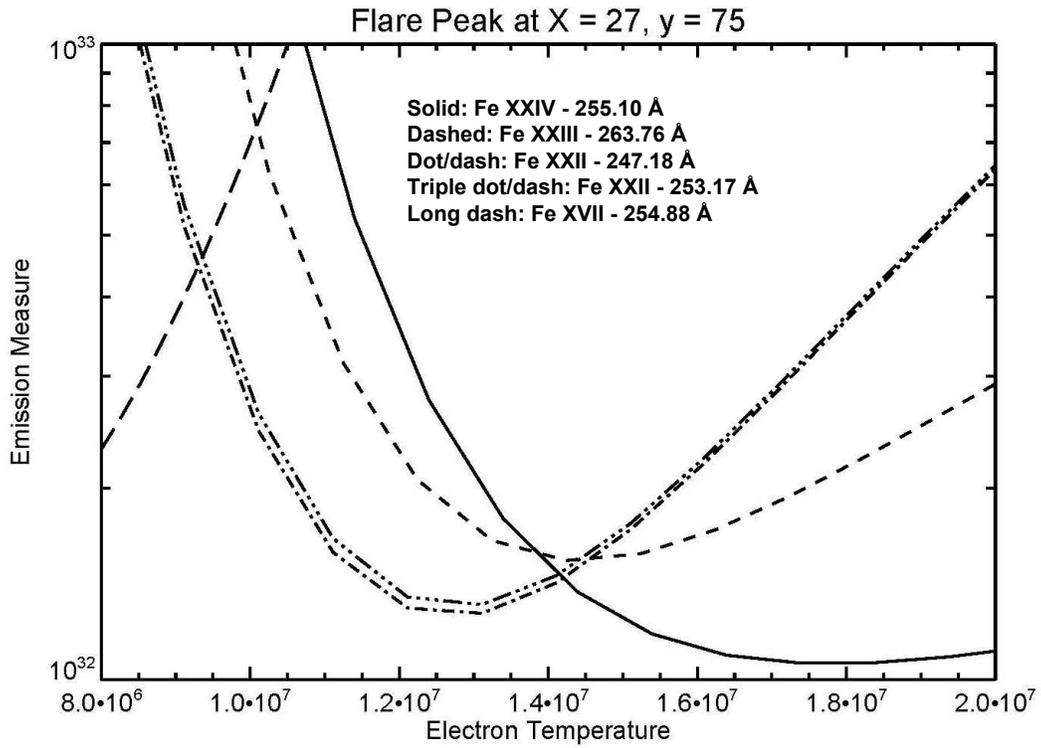}
\caption{\label{f17.eps} Loci plots for the lines shown assuming
  ionization equilibrium.  Except for the low-temperature \ion[Fe
  xvii] line, the curves cross near a common point and indicate an
  isothermal source at a temperature of about 13.7 MK.}
\end{figure}

3. In the coronal outflow regions the \ion[Fe xxiii] emission can be
modeled well with a single Gaussian, although a blueshifted component
may also be present, possibly due to line-of-sight effects.  The
single Gaussians are Doppler shifted to the blue by varying amounts
and the wavelengths (speeds) become larger (smaller) and approach the
rest wavelength near X = 27 (03:26:27 UT).  This implies that between
X = 5 (03:38:04 UT) and X = 15 (03:32:47 UT) we are observing a
plethora of loop footpoint regions with varying degrees of outflows
from two-arcsec pixel to pixel as is seen from Figure~\ref{f8.eps} and
Figure~\ref{f9.eps}.  The fact that the \ion[Fe xxiii] emission shows
net outflows contradicts the results of Milligan \& Dennis (2009) who
found a stationary component in addition to an outflow component at
footpoint regions and were puzzled by it since models predict net
shifts there.  However, the implication is that the actual strengths
of outflows and the temperatures at which they occur depend critically
on the energy input into the chromosphere from the corona, and perhaps
also on the nature of the input, i.e., conduction fronts or energetic
particles.  Observations can easily be complicated by line-of-sight
effects or unresolved footpoint regions.  Milligan \& Dennis (2009)
found downflows in the footpoint regions for coronal and transition
region lines, which also contradicts the present results.  The 9 March
flare is a considerably more energetic event than the event observed
by Milligan \& Dennis (2009), which probably explains the differences
in results.  We note that Kunichika Aoki \& Hirohisa Hara (oral paper,
Hinode 6 meeting, St. Andrews, Scotland 2012) also reported entire
shifts of the \ion[Fe xxiii] line centroid at footpoint regions from
EIS picket-fence raster flare studies.  Similarly, Brosius (2013)
obtained entire shifts of the \ion[Fe xxiii] profile from EIS spectra
obtained from a rapid cadence stare mode observation for a C1 flare
(See also Del Zanna et al. 2011).  Previously, Watanabe et al. (2010)
reported on variably shaped \ion[Fe xxiii] profiles from EIS spectra
in a flare and Young, Doschek, \& Warren (2013) also find \ion[Fe
  xxiii] profiles with intense blueshifted components.  The conclusion
is that highly variable results at footpoint regions can be expected
depending on energy input and probably also the impulsiveness of the
flare.  We have given results (Figure~\ref{f12.eps},
Figure~\ref{f13.eps}, and Figure~\ref{f14.eps}) for one X position for
a quantitative modeling exercise.

4. The region near (27,75) is special because of the strong energy
input that occurred there and produced the first flare peak, and the
presence of a sunquake close to the EIS slit position.  The redshifted
wing asymmetries that occur on the coronal lines are unique to the
region around (27,75).  The profiles of the multi-million degree flare
lines are well-fit by single Gaussians near or at the rest
wavelengths.  In the Standard Flare Model, this is the expectation for
the emission near the tops of loops after it has come to rest after
evaporation.  Such plasma could exist at this loop location because
the flare began about a minute prior to EIS observations.  In this
time plasma could have reached loop tops and produced stationary
emission.  Nevertheless, this emission appears quite impulsive in
time, i.e., it appears as a relatively small spot seen in Figure 1
(the EIS slit goes through it) around positions (2x27,75).  If the
energy impulse at this time is due to a reconnection event above the
closed loop tops as in the Standard Model, then the downflowing
coronal plasma might be due to the downward shock wave produced in the
reconnection.  If so, to our knowledge this is the first observation
of the downward propagating shock.  However, this is still simply a
speculation.  The downward flowing coronal plasma might also somehow
be linked to the sunquake, but unfortunately we have only snapshot
observations where higher spatial resolution is also needed to resolve
individual structures.

In summary, the EIS observations in this paper, coupled with results
in other papers, reveal multiple complex footpoint regions that give
rise to different evaporation dynamics over small spatial scales.
This indicates a complex energy input from the corona, suggesting
modelling to see if the EIS results can be reproduced by different
energy inputs, e.g., electron beams, conduction fronts, with different
values for total energy input and different temporal behavior, etc.

Tthe EIS observations make a strong case for the Extreme Ultra-Violet
Spectroscopic Telescope (EUVST) proposed for the Solar-C mission (see
Doschek et al., Solar-C white paper, Heliophysics Decadal Survey).
Proper slit observations for flares requires higher spatial
resolution, much higher temporal resolution, and greater temperature
coverage in the transition region and chromosphere.  Also desirable
are filter telescope images that do not saturate at flare onset and
that have clear bandpasses for picking out the multi-million degree
flare emission from coronal and transition region emission.

{\it Hinode} is a Japanese mission developed and launched by
ISAS/JAXA, collaborating with NAOJ as domestic partner, and NASA (USA)
and STFC (UK) as international partners.  Scientific operation of the
{\it Hinode} mission is conducted by the {\it Hinode} science team
organized at ISAS/JAXA.  This team mainly consists of scientists from
institutes in the partner countries.  Support for the post-launch
operation is provided by JAXA and NAOJ, STFC, NASA, ESA (European
Space Agency), and NSC (Norway).  We are grateful to the {\it Hinode}
team for all their efforts in the design, build, and operation of the
mission.

The authors acknowledge support from the NASA {\it Hinode} program and
from ONR/NRL 6.1 basic research funds.

\clearpage

\begin{table}\label{table1}
\caption{Spectral Lines, Wavelengths, and Temperatures of Formation}
\begin{center}
\begin{tabular}{lll}
\hline
\hline
Ion & Wavelength (\AA) & Temperature (K)\\
\hline
\ion[He ii] & $256.32$ & $5.0\times10^4$\\
\ion[O v] & $248.46$ & $2.2\times10^5$\\
\ion[O vi] & $184.12$ & $2.8\times10^5$\\
\ion[Fe VIII] & $185.211$ & $4.0\times10^5$\\
\ion[Fe x] & $184.534$ & $1.1\times10^6$\\
\ion[Fe xii] & $192.394$ & $1.6\times10^6$\\
\ion[Fe xii] & $195.119$ & $1.6\times10^6$\\
\ion[Fe xiii] & $202.047$ & $1.8\times10^6$\\
\ion[Fe xiii] & $203.824$ & $1.8\times10^6$\\
\ion[Fe xiv] & $264.783$ & $2.0\times10^6$\\
\ion[Fe xiv] & $274.202$ & $2.0\times10^6$\\
\ion[Fe xv] & $284.160$ & $2.2\times10^6$\\
\ion[Fe xvi] & $262.967$ & $2.8\times10^6$\\
\ion[Fe xvii] & $254.88$ & $4\times10^6$\\
\ion[Fe xxii] & $253.17$ & $1.3\times10^7$\\
\ion[Fe xxiii] & $263.76$ & $1.4\times10^7$\\
\ion[Fe xxiv] & $255.10$ & $1.8\times10^7$\\
\hline
\end{tabular}
\end{center}
\end{table}

\clearpage

\clearpage

\clearpage

\clearpage

\clearpage

\clearpage

\clearpage

\clearpage

\clearpage

\clearpage

\clearpage

\clearpage

\clearpage

\clearpage

\clearpage

\clearpage

\clearpage

\end{document}